\documentclass[12pt,thmsa,oneside]{book}
\usepackage{amssymb}

\usepackage{tcibkpk}
\usepackage{sw20wspc}

\input{epsfig}
\input{tcilatex}
\begin{document}

\author{Clovis J. de Matos \\
2, Nieuwsteeg 2311 SB Leiden The Netherlands\\
e-mail: cdematos@club-internet.fr \and Robert E. Becker\thanks{%
Main contact} \\
131 Old County Road, \# 168 Windsor Locks, CT 06096 USA\\
e-mail: roberte.becker@cwix.com}
\title{Gravitomagnetic Flux Quantization in Superconductors and a Method for the
Experimental Detection of Gravitomagnetism in the Terrestrial Laboratory}
\date{13th July 1999}
\maketitle

\begin{abstract}
\QTR{UserInput}{It is extraordinarily difficult to detect the extremely weak
gravitomagnetic (GM) field of even as large a body as the earth. To detect
the GM field, the gravitational analog of an ordinary magnetic field, in a
modest terrestrial laboratory should be that much more difficult. Here we
show, however, that for certain superconductor configuration and topologies,
it should be possible to detect a measurable GM field in the terrestrial
laboratory, by using the properties of superconductors imposed by quantum
mechanical requirements. In particular, we show that the GM Flux should be
quantized in a superconductor with non-vanishing genus, just like the
ordinary magnetic flux. And this magnetically induced, quantized GM Flux,
for sufficiently high quantum number and favorable geometries, should be
detectable, and distinguishable from the effects produced by an ordinary
magnetic field.}
\end{abstract}

\tableofcontents

\chapter{Introduction}

One of the predictions of General Theory of Relativity, unveiled eighty
years ago, is the existence, in a linear approximation, of a gravitational
analog of the magnetic field, known as the gravitomagnetic (GM) or
prorotational field. In the form of the Lens-Thirring Effect, this aspect of
the gravitational field has been studied extensively since then. Indeed,
there are two large scale space missions, Gravity Probe-B and LAGEOS\ III,
planned for launch near the turn of the millenium, designed to detect the
exceedingly weak GM field produced by the rotation of the earth on its own
axis.

It is the very weakness of the GM field of the entire earth that makes it so
difficult, if not outright impossible, to measure the GM field in a
terrestrial laboratory. Indeed, it has required extraordinary engineering
ingenuity and precision to design gyroscopes sensitive enough to detect the $%
10^{-15}Hz$ precession induced by the earth at the orbit of Gravity Probe-B.
Likewise, artificially generating a detectable GM field in the laboratory is
clearly hopless, though one of us has elsewhere \cite{Becker} proposed a
variant of a magnetic resonance experiment, utilizing scanning probe
microscopies and nanotechnologies, to attempt this.

The present method for the experimental detection of gravitomagnetism
(earlier versions are \cite{Becker}, \cite{Beckart}), proposes to overcome
the problem of the weakness of the GM field by taking advantage of the
unique properties of macroscopically coherent systems such as
superconductors (SC). It is by now well known that the magnetic flux
threading a superconductor is quantized with a quantum number $n$. We will
show that the GM field enters superconducting flux relations in the same
manner as the magnetic field, and that the GM flux should also be quantized.
Furthermore, if the magnetic flux quantum number is fixed for a given
physical configuration, that same quantum number governs the GM flux quantum
as well. And this GM flux quantum, for sufficiently high $n$, should also be
macroscopically detectable. This experiment is designed to detect the GM
field represented by this quantized GM flux.

One obvious question that must be asked is ''if the GM field of even a body
as massive as earth is so tiny, how is it conceivable that a small SC can
generate a field large enough to be readily measurable in the laboratory?''
To address this, first note that in the presence of an external GM field
there is an additional term in the usual quantized magnetic fluxoid
relation. This GM contribution to the magnetic flux is correctly taken to be
negligible because the external GM field is so tiny.

Similarly, there is a magnetic term in the quantized GM flux relation. For
substantial magnetic fields, such as those that would result in a high
magnetic flux quantum $n$, this term is not small. Thus, a large GM\ flux
quantum may be detected due to a magnetically-induced GM field in the SC.
The exact physical mechanism responsible for this ''amplification'' is still
nebulous. One of the objectives of this experimental method is to help
elucidate such a mechanism. Initial steps in identifying possible
theoretical mechanisms are outlined below in section 4.

It should be emphasized that a GM field large enough to be detectable in a
terrestrial laboratory is predicted to occur only in macroscopic coherent
systems such as superconductors or superfluids. (Though the effect should
also occur in superfluids with a toroidal topology, the supercarriers in
superfluids are much more massive than the Cooper pairs in SC, leading to a
much smaller GM\ flux quantum than can be found in a SC).

Ultimately, the effect to be studied, as with so many unusual phenomenon,
owes its postulated existence to the primacy of quantum mechanics, and the
even more special properties of macroscopic, coherent systems.

Non-detection of the effect would actually raise equally important
theoretical questions, such as: ''If the Maxwell approximation to general
relativity is correct, why is there not a GM flux quantum just as there is
for the magnetic field?'' After all, both vector potentials appear
''symmetrically'' in the quantum mechanical canonical momentum. ''Could the $%
U\left( 1\right) $ gauge field and topological description of the GM field
in the Maxwell approximation be wrong, even if the Maxwell approximation to
the Einstein equations is correct? What is the fundamental topological
relationship between magnetism, GM, and rotation?''

\chapter{Weak Gravitational Fields \& Gravitomagnetism}

In Einstein's general theory of relativity there exist gravitational
analogues of the electric and magnetic fields. We are already familiar with
the analogy between the electric field of a charge and the gravitational
field of a mass. It is well known that this analogy breaks down almost
immediately, because there are two kinds of charge and one kind of mass and
because two particles with the same type of charge repel, whereas particles
with the same type of mass attract. Nevertheless, if we are cognizant of
this distinction, we can still apply the analogy and still obtain useful
results.

We will show that the Einstein field equations not only can be made to agree
with this well known analogy between the electric and the gravitational
field, but they can also give rise to a gravitational analog to the magnetic
field. Its name is the \textbf{Gravitomagnetic }field. This gravitomagnetic
field has the dimensions of $\left( rad\cdot s^{-1}\right) $ in the MKS unit
system (this unit system will be used in our discussion), and is closely
related to Coriolis-type forces which arise from the principle of general
relativity.

In deriving this analogy between some of the gravitational forces and the
static and induction fields of electromagnetism, the following assumptions
have been made:

\begin{enumerate}
\item  \label{Apro_1}The mass densities are positive.

\item  \label{Apro_2}All motions are much slower than the speed of light, so
that special relativity can be neglected. (Often special relativistic
effects will hide general relativistic effects).

\item  \label{Apro_3}The kinetic or potential energy of all the bodies being
considered is much smaller than their mass energy.

\item  \label{Apro_4}The gravitational fields are always weak enough so that
superposition is valid.

\item  \label{Apro_5}The distance between objects is not so large that we
have to take retardation into account.(This can be ignored when we have a
stationary problem where the fields have already been prescribed and are not
changing with time).
\end{enumerate}

The procedure for linearizing Einstein's field equations is included in all
texts on general relativity \cite{Forward}. We start with Einstein field
equation: 
\begin{equation}
R_{\alpha \beta }-\frac{1}{2}g_{\alpha \beta }R=\frac{8\pi G}{c^{4}}%
T_{\alpha \beta }  \label{Ein_field_equ}
\end{equation}
Due to assumptions \ref{Apro_2} and \ref{Apro_4}, the metric tensor can be
approximated by 
\begin{equation}
g_{\alpha \beta }\simeq \eta _{\alpha \beta }+h_{\alpha \beta }
\label{Aprox_metric}
\end{equation}
where greek indices $\alpha ,\beta =0,1,2,3$ and $\eta _{\alpha \beta
}=\left( +1,-1,-1,-1\right) $ is the flat spacetime metric tensor, and $%
\left| h_{\alpha \beta }\right| <<1$ is the pertubation to the flat metric.
All the rationale that follows is correct up to the first order in $\left|
h_{\alpha \beta }\right| $.We assume that the $h_{\mu \nu ,\alpha \beta }$
are infinitesimally small up to first order; so $R_{\alpha \beta }$ and $R$
are also correct to first order. Therefore, we can consider $g_{\alpha \beta
}=\eta _{\alpha \beta }$. Using this form of the metric, the Ricci tensor
can be calculated from the contraction of the Riemann tensor and the
curvature scalar can be calculated from the contraction of the Ricci tensor 
\begin{equation}
R_{\alpha \beta }\simeq -\frac{1}{2}\square h_{\alpha \beta }  \label{Ricci}
\end{equation}
\begin{equation}
R\simeq \eta ^{\alpha \beta }R_{\alpha \beta }=\frac{1}{2}\square h,
\label{Scal_curv}
\end{equation}
where in obtaining (\ref{Ricci}) and (\ref{Scal_curv}) we choose our
coordinate system so that we have the following ''gauge'' condition 
\begin{equation}
\left[ h_{\alpha \beta }-\frac{1}{2}\eta _{\alpha \beta }h\right] _{,\beta
}=0.  \label{gauge}
\end{equation}
If we substitute the Ricci tensor (\ref{Ricci}) and the curvature scalar (%
\ref{Scal_curv}) into Einstein's equations, we obtain 
\begin{equation}
-\frac{1}{2}\square h_{\alpha \beta }+\frac{1}{4}\eta _{\alpha \beta
}\square h=\frac{8\pi G}{c^{4}}T_{\alpha \beta }  \label{Ein_equation_h}
\end{equation}
We now define the gravitational potential as 
\begin{equation}
\overline{h}_{\alpha \beta }=h_{\alpha \beta }-\frac{1}{2}\eta _{\alpha
\beta }h;  \label{hbar}
\end{equation}
substituting equation (\ref{hbar}) into equation (\ref{Ein_equation_h}) and
rearranging, we get 
\begin{equation}
\square \overline{h}_{\alpha \beta }=-\frac{16\pi G}{c^{4}}T_{\alpha \beta }.
\label{lin_equ}
\end{equation}
If we write out the D'Alembertian operator, we have 
\begin{equation}
\frac{1}{c^{2}}\frac{\partial ^{2}}{\partial t^{2}}\overline{h}_{\alpha
\beta }-\triangle \overline{h}_{\alpha \beta }=-\frac{16\pi G}{c^{4}}%
T_{\alpha \beta }.  \label{lin_dev}
\end{equation}
This is the basic equation upon which the analogies between electromagnetism
and gravity are based.

When spacetime is highly dynamical, for example around two colliding black
holes, there is no natural, preferred way to split spacetime into space plus
time. This fact has driven relativists, beginning with Einstein, to describe
gravity in terms of a unified, four dimensional spacetime with dynamical
evolving four dimensional curvature.

On the other hand, astrophysicists and experimental physicists usually deal
with situations where spacetime is stationary rather than dynamical, for
example the spacetime around the earth, or around a quiescent black hole. In
such cases stationarity dictates a preferred way to slice spacetime into
three-dimensional space plus one-dimensional time \cite{Thorne}. Although
such ''3+1 split'' are rarely treated in standard textbooks on general
relativity, they are used widely by professional relativists in numerical
solutions of the Einstein field equations, in the quantization of general
relativity, in astrophysical studies of black holes, and in analysis of
laboratory experiments to test general relativity.

The 3+1 split regards three dimensional space as curved rather than
Euclidean; its metric $g_{ik}$ (in an appropriate coordinate system) is just
the spatial part of the spacetime metric $g_{\alpha \beta }$. In this curved
three space reside two gravitational potentials: a ''\textbf{gravitoelectric}%
'' scalar potential $\Phi $, which is essentialy the time-time part $g_{00}$
of the space-time metric; and a ''\textbf{gravitomagnetic}'' vector
potential $\overrightarrow{A}_{g}$, which is essentially the time-space part 
$g_{0j}$ of the space-time metric. The decomposition of $g_{\alpha \beta }$
into $g_{ik}$, $\Phi $ and $\overrightarrow{A}_{g}$ is analogous to the
decomposition of the four vector potential $A_{\alpha }$ into an electric
scalar potential $\Psi =A_{0}$ and a magnetic vector potential $%
\overrightarrow{A}=A_{j}$.

\section{Gravitational Scalar Potential}

In the first approximation (zero order for the energy momentum tensor), we
assume that all quantities are not varying with time. Then the time
derivative of the gravitational potential is zero and all components of the
energy-momentum tensor are zero except 
\begin{equation}
T_{00}=\mu c^{2}.  \label{clov1}
\end{equation}
Equation (\ref{lin_dev}) reduces to 
\begin{equation}
-\triangle \overline{h}_{00}=-\frac{16\pi G}{c^{2}}\mu ,  \label{scal}
\end{equation}
which is essentially the Poisson equation, which has the solution 
\begin{equation}
\overline{h}_{00}=-\frac{4G}{c^{2}}\stackunder{V}{\iiint }\frac{\mu }{r}dV.
\label{Ari1}
\end{equation}
If we define the gravitational permittivity of the vacuum as 
\begin{equation}
\gamma =\frac{1}{4\pi G},  \label{clov2}
\end{equation}
we get 
\begin{equation}
\frac{c^{2}\overline{h}_{00}}{4}=-\frac{1}{4\pi \gamma }\stackunder{V}{%
\iiint }\frac{\mu }{r}dV.  \label{scal_pot}
\end{equation}
Comparing equation (\ref{scal_pot}) with the scalar potential of an electric
charge density 
\begin{equation}
\varphi =-\frac{1}{4\pi \varepsilon _{0}}\stackunder{V}{\iiint }\frac{\rho }{%
r}dV  \label{Ari2}
\end{equation}
we see that we can construct the well known gravitational analog to the
scalar potential: 
\begin{equation}
\Phi =\frac{c^{2}\overline{h}_{00}}{4}=\frac{c^{2}\left( g_{00}-1\right) }{2}%
.  \label{final_grav_pot}
\end{equation}

\section{Space Curvature}

This first approximation (\ref{scal}) also determines the spatial metric.
The existence of the component $\overline{h}_{00}$ results in a relativistic
interval of the form of the Schwarschild metric (using a cartesian
coordinate system $x^{1}=x$, $x^{2}=y$, $x^{3}=z$) 
\begin{equation}
ds^{2}=\left( 1+\frac{2\Phi }{c^{2}}\right) c^{2}dt^{2}-\left( 1-\frac{2\Phi 
}{c^{2}}\right) \left( dx^{2}+dy^{2}+dz^{2}\right) .  \label{clov3}
\end{equation}
Thus, the three dimensional spatial metric will be of the form 
\begin{equation}
g_{ab}=\left[ 
\begin{array}{ccc}
-\left( 1-\frac{2\Phi }{c^{2}}\right) & 0 & 0 \\ 
0 & -\left( 1-\frac{2\Phi }{c^{2}}\right) & 0 \\ 
0 & 0 & -\left( 1-\frac{2\Phi }{c^{2}}\right)
\end{array}
\right] .  \label{clov4}
\end{equation}
In higher approximations, the additional terms in the spatial metric will be
smaller than $2\Phi /c^{2}$ by the order of $\left( v/c\right) ^{2}$, and
since we assume velocities much smaller than the speed of light, they will
be of little experimental interest here.

\section{Gravitomagnetic Vector Potential}

In the next higher approximation (first order approximation for the
energy-momentum tensor), we still assume that the potential is not varying
with time, but that the masses involved have appreciable velocity or
rotation. Then the energy-momentum tensor will have the components 
\begin{equation}
T_{00}=\mu c^{2}  \label{clov5}
\end{equation}
and 
\begin{equation}
T_{0i}=-\mu cv_{i}.  \label{En_comp}
\end{equation}
We then have four equations remaining: One gives us the scalar potential
obtained previously, and the other three are 
\begin{equation}
-\triangle \overline{h}_{0i}=\frac{16\pi G}{c^{3}}\mu v_{i}.  \label{clov6}
\end{equation}
This equation has the solution 
\begin{equation}
\overline{h}_{0i}=\frac{4G}{c^{3}}\stackunder{V}{\iiint }\frac{\mu v_{i}}{r}%
dV.  \label{vect}
\end{equation}
If we define a gravitational permeability of space by 
\begin{equation}
\eta =\frac{4\pi G}{c^{2}}  \label{perme}
\end{equation}
then we can substitute (\ref{perme}) into (\ref{vect}) and rearrange to get 
\begin{equation}
-\frac{c\overline{h}_{0i}}{4}=-\frac{\eta }{4\pi }\stackunder{V}{\iiint }%
\frac{\mu v_{i}}{r}dV.  \label{clov7}
\end{equation}
Thus, we can identify a mass density flow $\overrightarrow{p}=\mu 
\overrightarrow{v}$ as a gravitational equivalent to the magnetic vector
potential whose components are the three components 
\begin{equation}
A_{g_{i}}=-\frac{c}{4}\overline{h}_{0i}=-\frac{c}{4}g_{0i}
\label{Vector_potential}
\end{equation}
(the factor of 4 appearing in the denominator of equation (\ref
{Vector_potential}) will be explained when we will reach equation (\ref{U_gm}%
))and thereby arrive at the isomorphism of the equations \cite{Ciufolini} 
\begin{equation}
\overrightarrow{A}_{g_{i}}=-\frac{\eta }{4\pi }\stackunder{V}{\iiint }\frac{%
\overrightarrow{p}}{r}dV\qquad and\qquad \overrightarrow{A}=+\frac{\mu _{0}}{%
4\pi }\stackunder{V}{\iiint }\frac{\overrightarrow{j}}{r}dV.  \label{clov8}
\end{equation}

\section{The Einstein-Maxwell-Type Gravitational Equations}

Let us change the linearized Einstein field equations into the form of
Maxwell Equations \cite{Peng} 
\begin{equation}
-\frac{1}{2}\left( \overline{h}_{\alpha \beta ,\mu }^{,\mu }+\eta _{\alpha
\beta }\overline{h}_{\mu \nu }^{,\mu \nu }-\overline{h}_{\alpha \mu ,\beta
}^{,\mu }-\overline{h}_{\beta \mu ,\alpha }^{,\mu }\right) =\frac{8\pi G}{%
c^{4}}T_{\alpha \beta }  \label{clov9}
\end{equation}
\begin{equation}
\frac{1}{4}\frac{\partial }{\partial x^{\mu }}\left( \overline{h}_{\alpha
\beta ,\mu }-\overline{h}_{\alpha \mu ,\beta }+\eta _{\alpha \beta }%
\overline{h}_{\mu \nu }^{,\nu }-\eta _{\alpha \mu }\overline{h}_{\beta \nu
}^{,\nu }\right) =-\frac{4\pi g}{c^{4}}T_{\alpha \beta }.  \label{clov10}
\end{equation}
For the sake of convenience we have introduced the tensor 
\begin{equation}
G_{\alpha \beta \mu }=\frac{1}{4}\left( \overline{h}_{\alpha \beta ,\mu }-%
\overline{h}_{\alpha \mu ,\beta }+\eta _{\alpha \beta }\overline{h}_{\mu \nu
}^{,\nu }-\eta _{\alpha \mu }\overline{h}_{\beta \nu }^{,\nu }\right) .
\label{clov11}
\end{equation}
Due to the ''gauge condition'' (\ref{gauge}) 
\begin{equation}
\overline{h}_{\mu \nu }^{,\nu }=0  \label{clov12}
\end{equation}
the tensor $G_{\alpha \beta \mu }$ can be simplified to 
\begin{equation}
G_{\alpha \beta \mu }=\frac{1}{4}\left( \overline{h}_{\alpha \beta ,\mu }-%
\overline{h}_{\alpha \mu ,\beta }\right) .  \label{Tensor_G}
\end{equation}
This tensor has the following properties 
\begin{equation}
G^{\alpha \beta \mu }=-G^{\alpha \mu \beta }  \label{clov13}
\end{equation}
\begin{equation}
G^{\alpha \beta \mu }+G^{\mu \alpha \beta }+G^{\beta \mu \alpha }=0
\label{clov14}
\end{equation}
\begin{equation}
G^{\alpha \beta \mu ,\lambda }+G^{\alpha \lambda \beta ,\mu }+G^{\alpha \mu
\lambda ,\beta }=0.  \label{Max1}
\end{equation}
With the help of the tensor $G_{\alpha \beta \mu }$ the linearized Einstein
field equations (\ref{lin_equ}) become 
\begin{equation}
\frac{\partial G^{\alpha \beta \mu }}{\partial x^{\mu }}=-\frac{4\pi G}{c^{4}%
}T^{\alpha \beta }.  \label{Max2}
\end{equation}
Introducing the gravitational scalar potential and the GM vector potential
we obtain equations (\ref{final_grav_pot}) and (\ref{Vector_potential}),
respectively 
\begin{equation}
\Phi =+\frac{c^{2}\overline{h}^{00}}{4}  \label{g1}
\end{equation}
\begin{equation}
A_{g}^{i}=-\frac{c}{4}\overline{h}^{0i}\qquad \overrightarrow{A}_{g}=\left(
A_{g}^{1},A_{g}^{2},A_{g}^{3}\right) .  \label{g2}
\end{equation}
Introduce new signs, and substitute equation (\ref{g1}) into equation (\ref
{Tensor_G}), to get the gravitoelectric field 
\begin{equation}
G^{00i}=\frac{1}{4}\left( \overline{h}^{00,i}-\overline{h}^{0i,0}\right) 
\label{clov15}
\end{equation}
\begin{equation}
g^{i}=-c^{2}G^{00i}=-\frac{\partial \Phi }{\partial x^{i}}-\frac{\partial
A_{g_{i}}}{\partial t}  \label{clov16}
\end{equation}
\begin{equation}
\overrightarrow{g}=-\nabla \Phi -\frac{\partial \overrightarrow{A}_{g}}{%
\partial t}.  \label{Maxg}
\end{equation}
Substitute equation (\ref{g2}) into equation (\ref{Tensor_G}) to get the
gravitomagnetic field 
\begin{equation}
cG^{0ij}=A_{g}^{i,j}-A_{g}^{j,i}  \label{clov17}
\end{equation}
or 
\begin{equation}
\overrightarrow{B}_{g}=\nabla \wedge \overrightarrow{A}_{g}.  \label{Maxbgm}
\end{equation}
Using the expressions for the gravitoelectric and GM fields given as
functions of tensor $\overline{h}^{\alpha \beta }$ derivatives, we can
obtain from equation (\ref{Max2}) the ''Gauss law for the gravitoelectric
field'' and the ''gravitational Ampere law''; and from equation (\ref{Max1})
we get the ''Gauss law for the gravitomagnetic field'' and the ''Faraday
induction law for the gravitoelectric field''. These four laws constitute
the four Einstein-Maxwell-type gravitational equations, i.e., 
\begin{equation}
\nabla \overrightarrow{g}=-4\pi G\mu   \label{clov18}
\end{equation}
\begin{equation}
\nabla \overrightarrow{B}_{g}=0  \label{clov19}
\end{equation}
\begin{equation}
\nabla \wedge \overrightarrow{g}=-\frac{\partial \overrightarrow{B}_{g}}{%
\partial t}  \label{clov20}
\end{equation}
\begin{equation}
\nabla \wedge \overrightarrow{B}_{g}=-\frac{4\pi G}{c^{2}}\mu 
\overrightarrow{v}+\frac{1}{c^{2}}\frac{\partial \overrightarrow{g}}{%
\partial t}.  \label{clov21}
\end{equation}

\section{The Equations of Motion in the Weak Field Approximation}

In the four-dimensional equation of motion 
\begin{equation}
\frac{du^{\lambda }}{d\tau }+\Gamma _{\mu \nu }^{\lambda }u^{\mu }u^{\nu }=0,
\label{clov23}
\end{equation}
the gravitational effects are entirely in the metric tensor. The only forces
explicitly stated are nongravitational forces. This four-dimensional
equation of motion can be broken down and arranged so that it is a
three-dimensional curvilinear spatial equation of motion. The gravitational
effects resulting from the temporal components of the metric tensor are
represented as forces due to a gravitational scalar and a gravitational
vector potential. The spatial components of the metric tensor are used as
the three-dimensional metric tensor.

The general equation of motion for a particle with only gravitational forces
acting is given by Moller \cite{Moller} as 
\begin{equation}
\frac{dP_{\alpha }}{d\tau }-\frac{1}{2}\frac{\partial g_{\mu \nu }}{\partial
x^{\alpha }}U^{\mu }P^{\nu }=0,  \label{geo}
\end{equation}
where $d\tau =ds/c$, 
\begin{equation}
P_{\alpha }=mU_{\alpha }=m_{0}g_{\alpha \lambda }U^{\lambda }=m_{0}g_{\alpha
\lambda }\Gamma \frac{dx^{\lambda }}{dt},  \label{clov22}
\end{equation}
and 
\begin{equation}
\Gamma =\frac{dt}{d\tau }=\left[ 1+\frac{2\Phi }{c^{2}}-\left( \frac{v}{c}%
\right) ^{2}-\frac{8}{c^{2}}A_{g_{i}}v^{i}\right] ^{-1/2}.  \label{gam}
\end{equation}
Note that if $\Phi =\overrightarrow{A}_{g}=0$ then we recover the usual
relativistic correction: 
\begin{equation}
\Gamma =\frac{1}{\sqrt{1-\left( \frac{v}{c}\right) ^{2}}}.  \label{clov24}
\end{equation}
Rearranging equation (\ref{geo}) we have 
\begin{equation}
\frac{1}{\Gamma }\frac{d}{dt}\left( g_{\alpha \lambda }\Gamma \frac{%
dx^{\lambda }}{dt}\right) =\frac{1}{2}\frac{\partial g_{\mu \nu }}{\partial
x^{\alpha }}\frac{dx^{\mu }}{dt}\frac{dx^{\nu }}{dt}  \label{motion}
\end{equation}

The $\alpha =0$ equation gives us the conservation of mass-energy: 
\begin{equation}
\frac{1}{\Gamma }\frac{d}{dt}\left[ \Gamma \left( g_{00}\frac{dx_{0}}{dt}%
+g_{0i}\frac{dx^{i}}{dt}\right) \right] =\frac{1}{2}\frac{\partial g_{ii}}{%
\partial x^{0}}\left( \frac{dx^{i}}{dt}\right) ^{2}+\frac{1}{2}\frac{%
\partial g_{00}}{\partial x^{0}}\left( \frac{dx^{0}}{dt}\right) ^{2}+\frac{%
\partial g_{0i}}{\partial x^{0}}\left( \frac{dx^{0}}{dt}\frac{dx^{i}}{dt}%
\right) .  \label{Ari3}
\end{equation}
Letting $dx^{0}/dt=c$ and considering $\Gamma \simeq 1\,$and $g_{ii}=\left(
-1,-1,-1\right) $, we get: 
\begin{equation}
\frac{d}{dt}\left( cg_{00}+g_{0i}\frac{dx^{i}}{dt}\right) =\frac{c}{2}\frac{%
\partial g_{00}}{\partial t}+\frac{\partial g_{0i}}{\partial t}\frac{dx^{i}}{%
dt}.  \label{Ari4}
\end{equation}
We then use our definition of the gravitational and vector potential 
\begin{equation}
g_{00}=1+\frac{2\Phi }{c^{2}}\qquad and\qquad g_{0i}=-\frac{4}{c}A_{g_{i}}
\label{Ari5}
\end{equation}
to get 
\begin{equation}
\frac{d}{dt}\left( \Phi -4A_{g_{i}}v^{i}\right) =g_{i}v^{i}  \label{Ari6}
\end{equation}
\begin{equation}
-\frac{d\left( m\Phi \right) }{dt}+\frac{d\left( 4m\overrightarrow{v}\cdot 
\overrightarrow{A}_{g}\right) }{dt}=m\overrightarrow{g}\cdot \overrightarrow{%
v}.  \label{Energy_balance}
\end{equation}
From equation (\ref{Energy_balance}) we conclude that the gravitational
potential energy and the gravitomagnetic potential energy are respectively
given by: 
\begin{equation}
U_{grav}=m\Phi  \label{U_grav}
\end{equation}
and 
\begin{equation}
U_{gm}=-\stackrel{\downarrow }{4}m\overrightarrow{v}\cdot \overrightarrow{A}%
_{g}.  \label{U_gm}
\end{equation}

The factor 4 appearing in equations (\ref{U_gm}), (\ref{Vector_potential}), (%
\ref{det}) and (\ref{equmo}), is presumably due to gravity being associated
with a spin-2 (graviton) field rather than a spin-1 field (photon).

The other three equations for $\alpha =i=1,2,3$ equation (\ref{motion}) are
the equations of motion\cite{Davidson}. 
\begin{equation}
\frac{1}{\Gamma }\frac{d}{dt}\left[ \Gamma \left( g_{i0}\frac{dx^{0}}{dt}%
+g_{ij}\frac{dx^{j}}{dt}\right) \right] =\frac{1}{2}\frac{\partial g_{00}}{%
\partial x^{i}}\left( \frac{dx^{0}}{dt}\right) ^{2}+\frac{\partial g_{0j}}{%
\partial x^{i}}\left( \frac{dx^{0}}{dt}\frac{dx^{j}}{dt}\right) +\frac{1}{2}%
\frac{\partial g_{kj}}{\partial x^{i}}\left( \frac{dx^{k}}{dt}\frac{dx^{j}}{%
dt}\right) .  \label{Ari7}
\end{equation}
Putting $dx^{0}/dt=c$ and considering $\Gamma \simeq 1\,$and $g_{ii}=\left(
-1,-1,-1\right) $, we get: 
\begin{equation}
\frac{d}{dt}\left( g_{ij}\frac{dx^{j}}{dt}\right) =\frac{c^{2}}{2}\frac{%
\partial g_{00}}{\partial x^{i}}-c\frac{\partial g_{i0}}{\partial t}+c\left( 
\frac{\partial g_{0j}}{\partial x^{i}}-\frac{\partial g_{0i}}{\partial x^{j}}%
\right) \frac{dx^{j}}{dt}.  \label{Ari8}
\end{equation}
We then use our definition of the gravitational and vector potential 
\begin{equation}
g_{00}=1+\frac{2\Phi }{c^{2}}\qquad and\qquad g_{0i}=-\frac{4}{c}A_{g_{i}}
\label{Ari9}
\end{equation}
to get: 
\begin{equation}
\frac{d}{dt}\left( g_{ij}\frac{dx^{j}}{dt}\right) =\frac{\partial \Phi }{%
\partial x^{i}}+4\frac{\partial \overrightarrow{A}_{g_{i}}}{\partial t}%
-4\left( \overrightarrow{v}\wedge \overrightarrow{B}_{g}\right) _{i}.
\label{Ari10}
\end{equation}
Rearranging we have: 
\begin{equation}
\frac{d^{2}x^{i}}{dt^{2}}=\left( -\frac{\partial \Phi }{\partial x^{i}}-%
\stackrel{\downarrow }{4}\frac{\partial \overrightarrow{A}_{g_{i}}}{\partial
t}\right) +\stackrel{\downarrow }{4}\left( \overrightarrow{v}\wedge 
\overrightarrow{B}_{g}\right) _{i}.  \label{det}
\end{equation}
Using equation \ref{Maxg}, we obtain
\begin{equation}
m\overrightarrow{a}=m\left( \overrightarrow{g}-3\frac{\partial 
\overrightarrow{A}_{g}}{\partial t}\right) +4m\overrightarrow{v}\wedge 
\overrightarrow{B}_{g}.  \label{equmo}
\end{equation}
The left hand side of equation (\ref{det}) is the total acceleration of the
particle. The first term on the right hand side is one component of $\nabla
\Phi $, the gravitational static attraction; the second term is one
component of $\partial \overrightarrow{A}_{g}/\partial t$, the gravitational
induction effect; and the third term is one component of $\overrightarrow{v}%
\wedge \left( \nabla \wedge \overrightarrow{A}_{g}\right) $ the
gravitational equivalent of the Lorentz force.

\section{From Electromagnetism to Gravity}

When the analogy between electrodynamics and gravity is carried out and all
the constants are evaluated, we obtain an isomorphism between the
gravitational and the electromagnetic quantities.

\begin{center}
\begin{tabular}{|p{1.15in}|p{0.5in}|p{0.8in}|p{2.2in}|}
\hline
& EM Symbol & Gravitational Symbol & Value or definition \\ \hline
Force Vector & $\overrightarrow{E}$ & $\overrightarrow{g}$ & $=-\nabla \Phi -%
\frac{\partial \overrightarrow{A}_{g}}{\partial t}$ \\ \hline
Solenoidal Force Vector & $\overrightarrow{B}$ & $\overrightarrow{B}_{g}$ & $%
=\nabla \wedge \overrightarrow{A}_{g}$ \\ \hline
Scalar Potential & $\varphi $ & $\Phi $ & $\simeq -\frac{1}{4\pi \gamma }%
\stackunder{V}{\iiint }\frac{\mu }{r}dV$ \\ \hline
Vector Potential & $\overrightarrow{A}$ & $\overrightarrow{A}_{g}$ & $\simeq
-\frac{\eta }{4\pi }\stackunder{V}{\iiint }\frac{\mu \overrightarrow{v}}{r}%
dV $ \\ \hline
Source Density & $\rho $ & $\mu $ & $=\frac{dM}{dV}$ \\ \hline
Source Quantity & $Q$ & $M$ & $=\stackunder{V}{\iiint }\mu dV$ \\ \hline
Current Density & $\overrightarrow{j}$ & $\overrightarrow{p}$ & $=\mu 
\overrightarrow{v}$ \\ \hline
Current quantity & $I$ & $\stackrel{\cdot }{M}$ & $=\frac{dM}{dt}=%
\stackunder{S}{\iint }\overrightarrow{p}\cdot \widehat{n}dS$ \\ \hline
Moment & $\overrightarrow{M}$ & $\overrightarrow{J}=\frac{1}{2}%
\overrightarrow{L}$ & $=\frac{1}{2}I\omega $ \\ \hline
Permittivity of Space & $\varepsilon $ & $\gamma $ & $=\frac{1}{4\pi G}%
=1.19\times 10^{9}\left[ kg.s^{2}/m^{3}\right] $ \\ \hline
Permeability of Space & $\mu $ & $\eta $ & $=\frac{4\pi G}{c^{2}}=9.33\times
10^{-27}\left[ m/kg\right] $ \\ \hline
\end{tabular}

Table 1: From electromagnetism to gravity
\end{center}

The ordinary electromagnetic (EM) magnetic flux density $\overrightarrow{B}$
can be defined by its torque on a magnetic moment 
\begin{equation}
\overrightarrow{N}=\overrightarrow{M}\wedge \overrightarrow{B}  \label{mot}
\end{equation}
where $\overrightarrow{N}$ is the torque and $\overrightarrow{M}$ the
magnetic moment. Since torque is defined to be the time rate of change of
angular momentum $\overrightarrow{L}$, and since $\overrightarrow{M}$ is
related to $\overrightarrow{L}$ in many classical systems by 
\begin{equation}
\overrightarrow{M}=\frac{\overline{q}}{2m}\overrightarrow{L},  \label{clov25}
\end{equation}
where $\overline{q}$ is the charge and $m$ the mass of the particle, we can
rewrite (\ref{mot}) as 
\begin{equation}
\frac{d\overrightarrow{L}}{dt}=\frac{\overline{q}}{2m}\overrightarrow{L}%
\wedge \overrightarrow{B}.  \label{dt}
\end{equation}
It is well known that (\ref{dt}) represents the precession of the angular
momentum $\overrightarrow{L}$ about the direction of $\overrightarrow{B}$ at
a Larmor frequency of 
\begin{equation}
\overrightarrow{\omega }=\frac{\overline{q}}{2m}\overrightarrow{B}.
\label{clov26}
\end{equation}

Just as a spinning magnetic moment generates a magnetic field, solutions of
the Einstein Equations show that a spinning mass generates a dipolar GM
field $\overrightarrow{B}_{g}$. In this case, the GM equivalent of the
magnetic moment is the angular momentum $\overrightarrow{L}$, so that the GM
torque on a spinning body can be written as 
\begin{equation}
\overrightarrow{N}=2\overrightarrow{L}\wedge \overrightarrow{B}_{g}.
\label{elle}
\end{equation}
Comparing (\ref{elle}) with (\ref{dt}) we conclude 
\begin{equation}
\frac{d\overrightarrow{L}}{dt}=2\overrightarrow{L}\wedge \overrightarrow{B}%
_{g}.  \label{clov27}
\end{equation}
Thus, there is a GM precessional frequency of 
\begin{equation}
\overrightarrow{\omega }_{g}=2\overrightarrow{B}_{g}.  \label{clov28}
\end{equation}
It is $\overrightarrow{\omega }_{g}$ that we will attempt to detect in our
experiment.

It should be emphasized that the previous discussion is approximate and is
presented merely to provide a simple tool with which to make estimates.

\chapter{Superconductivity}

Superconductors (SC) have many curious characteristics \cite{Tilley}, \cite
{Rose}. They are perfect conductors, so that once a current is started for
any reason, it will continue. But they also obey the Meissner effect by
which a magnetic field is excluded from the interior of a bulk SC, even in
an applied $\overrightarrow{B}\,$field. Physically, this comes about because
the $\overrightarrow{B}$ field induces surface (not bulk) currents, which
set up a field opposing the applied field in the interior; the net field is $%
\overrightarrow{B}=0$. These currents only flow in a layer, the London
penetration depth, in which $\overrightarrow{B}$ and the current $%
\overrightarrow{j}$ fall-off exponentially with distance into the SC.

All the strange properties of SC ultimately derive from the quantum
mechanical (QM) state of the system. Below a certain critical temperature,
it becomes energetically favorable for conduction electrons in the material
to form pairs such that their center of mass momentum is zero; these are
Cooper pairs. At absolute zero, all the electrons in the SC form Cooper
pairs, and so condense into the QM state of zero momentum.

Since electrons have a QM spin of $1/2$, and since opposite spin electrons
are paired, a Cooper pair is a QM object of integral spin, called a boson.
Unlike fermions, which obey the Pauli Exclusion Principle, any number of
bosons can occupy the same QM state. So below the critical temperature, $%
T_{c}$, the SC can be said to be in a macroscopically occupied QM state. In
order to change the state, a physical phenomena would have to break all the
Cooper pairs and remove the electrons from this state, known as a
condensate. This requires much more energy than just breaking a single
Cooper pair, and so changing QM states becomes exceedingly unlikely.

This explains properties like persistent currents. They persist, because
once the QM state is set-up, it can not easily be changed. There are
complications to the simple picture just described. For example, persistant
London screening currents must be generated by an applied field, and then
the Cooper pairs do not have zero center of mass momentum. But they do have
zero QM canonical momentum, which is the correct, vanishing quantity in this
case.

At $T>0$, not all the electrons form Cooper pairs, but the ground state is
still macroscopically occupied. For our purposes, this will suffice. Another
complication is the existence of several types of SC. The first to be
discovered was Type I SC, which obey the Meissner effect for all $B$ and all 
$T<T_{c}$. Next came Type II SC, which obey Meissner effect only for $%
B<B_{c_{1}}$, the lower Critical Field. Between $B_{c_{1}}$ and a $B_{c_{2}}$
(Upper Critical Field), an applied magnetic field can penetrate the SC and
form current vortices around the field lines.

Finally, there are the famous high-$T_{c}$, SC. High-$T_{c}$ SC do not obey
the same microscopic QM theory as do standard Type I and Type II\ SC (s-wave
BCS theory), but for our purposes can be regarded as Type II SC: they have a 
$B_{c_{1}}$. In the non-Meissner regime of the Type II SC \cite{Tilley}, the
critical property is that the magnetic field can penetrate in a vortex. The
vortex core is not SC, and is composed of normal material; that is,
electrons not paired into Cooper pairs. Topologically, this has enormous
importance, because it means the SC material is no longer simply-connected;
it now has a hole in it. The same thing can happen to any SC if one drills a
hole into an otherwise bulk SC, creating a ring or doughnut shaped object.

When we regard the whole SC as being in a condensate state, we can decribe
it by a single, condensate wave function and the Schr\"{o}dinger equation
for a Cooper pair will be similar to the following equation \cite{Feynmann}: 
\begin{equation}
-\frac{\hbar }{i}\frac{\partial \psi }{\partial t}=\widehat{H}\psi =\frac{1}{%
2m}\left( \frac{\hbar }{i}\nabla -q\overrightarrow{A}\right) ^{2}\psi
+q\varphi \psi .  \label{schro}
\end{equation}
This is the Schr\"{o}dinger equation for a particle with electric charge $q$
and mass $m$ moving in an electromagnetic field. In equation (\ref{schro}), $%
\varphi $ is the electrostatic potential and $\overrightarrow{A}$ is the
magnetic vector potential. Note that $\varphi $ and $\overrightarrow{A}$ are
applied potentials on, not the field generated by, the moving charge, $q$.

Notice that in the case of the Cooper pairs the charge $q$ and the mass $m$
that appear in equation (\ref{schro}) are respectively, twice the electron's
charge and twice the electron's mass.

As we said above, the Cooper pairs are bosons, therefore the amplitude to
find almost all the Cooper pairs present in a SC in the same QM state is
very large. In an ideal, non-interacting system, the pairs will be in the
state of lowest energy, the ground state.

Let $\psi $ be the wave function of a Cooper pair in the state of lowest
energy. The probability to find the pair in this state will be proportional
to the electric charge density $\rho $%
\begin{equation}
\psi \psi ^{*}\varpropto \rho .  \label{proba}
\end{equation}
Therefore, we can write the wave function in the following form: 
\begin{equation}
\psi \left( \overrightarrow{r}\right) =\sqrt{\rho \left( \overrightarrow{r}%
\right) }e^{i\theta \left( \overrightarrow{r}\right) },  \label{wavef}
\end{equation}
where $\rho $ and $\theta $ are real functions of $\overrightarrow{r}$. What
is the physical meaning of the phase $\theta $ in the wave function?

To answer this question we have to write down the current density for a
charged particle moving in an EM field 
\begin{equation}
\overrightarrow{j}=\frac{1}{2}\left\{ \left[ \frac{\widehat{P}-q%
\overrightarrow{A}}{m}\psi \right] ^{*}\psi +\psi ^{*}\left[ \frac{\widehat{P%
}-q\overrightarrow{A}}{m}\psi \right] \right\}  \label{curent}
\end{equation}
where $\widehat{P}=\frac{\hbar }{i}\nabla $ is the linear momentum operator.
Putting equation (\ref{wavef}) into (\ref{curent}), we have the following
equation for the current: 
\begin{equation}
\overrightarrow{j}=\frac{\hbar }{m}\left( \nabla \theta -\frac{q}{\hbar }%
\overrightarrow{A}\right) \rho .  \label{cur}
\end{equation}
Inserting $\overrightarrow{j}=\rho \overrightarrow{v}$ into equation (\ref
{cur}) we obtain: 
\begin{equation}
\hbar \nabla \theta =m\overrightarrow{v}+q\overrightarrow{A}.  \label{canon}
\end{equation}
Therefore, the canonical momentum $\overrightarrow{P}$ for a Cooper pair is
given by: 
\begin{equation}
\overrightarrow{P}=\hbar \nabla \theta .  \label{P}
\end{equation}
As $\overrightarrow{j}\,$and $\rho $ have a direct physical meaning, $\rho $
and $\theta $ are real quantities.\pagebreak

\section{Quantization of the Magnetic Flux}

Let us consider a conducting ring in a magnetic field.


\begin{figure}[hbtp]
\begin{center}
\mbox{\psfig{file=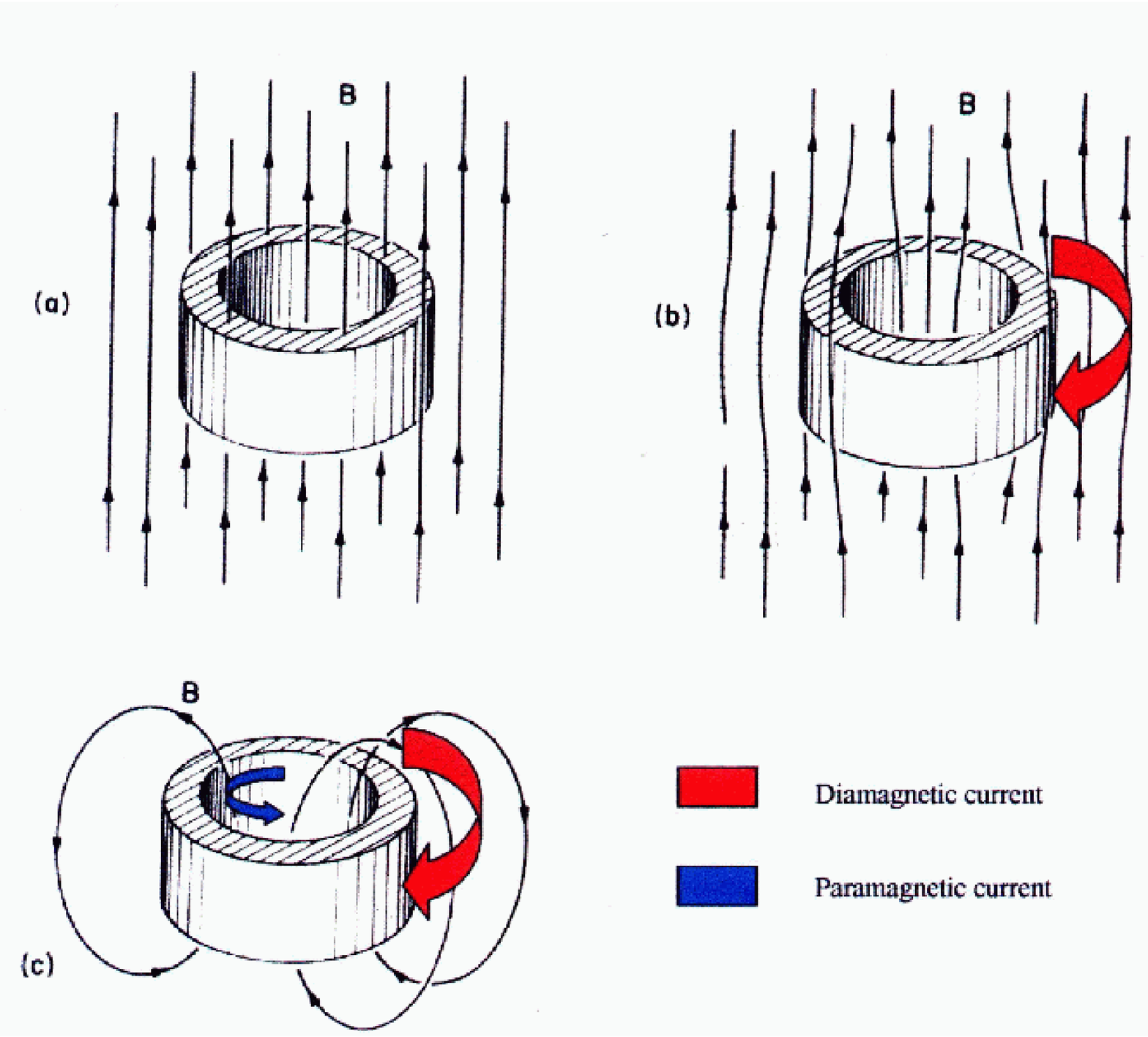,height=3.68in}}
\end{center}
\begin{center}Figure 1: a ring in an external
magnetic field, (a) in the normal state, (b) in the superconductive state,
(c) when the external magnetic field has been removed.
\end{center}
\end{figure}


In the normal state there is a magnetic field inside the ring (figure
1-(a)). When the ring becomes superconducting (at the critical temperature $%
T_{c}$), the magnetic field is expelled outside the material. At this moment
there will be a given magnetic flux through the ring's hole (figure 1-(b)).
If we now remove the external applied magnetic field, the lines of the
magnetic field that cross the ring's hole will be trapped (figure 1-(c)).
The magnetic flux $\Phi _{m}$ inside the central hole can not decrease,
because $\frac{\partial \Phi _{m}}{\partial t}$ must be equal to the line
integral of $\overrightarrow{E}$ along a closed path inside the ring. This
integral is zero inside a superconductor. When we remove the external field,
a circular supercurrent is established through the surface of the ring to
maintain the magnetic flux constant through the ring's hole.

Whereas screening, diamagnetic, currents are always set-up around the outer
periphery of a SC to cancel an applied field in the interior (see figure
1-(b)), if a hole is present, another current, called paramagnetic current 
\cite{Rose}, is set-up around the inner circumference of the hole (see
figure 1-(c)). The paramagnetic current ensures that once a SC is cooled
below $T_{c}$ in an applied field, the magnetic flux penetrating the hole
remains constant even if the applied field is removed.

Inside the ring the current density is zero. Therefore from the current
equation (\ref{cur}) we have: 
\begin{equation}
\hbar \nabla \theta =q\overrightarrow{A}.  \label{clov29}
\end{equation}
Now we take the integral of $\overrightarrow{A}$ over a closed path $\Gamma $
(see figure 2) that goes around the ring hole and inside the central region
of the SC bulk, in a way that it will never get close to the surface of the
ring.


\begin{figure}[htb]
\begin{center}
\mbox{\psfig{file=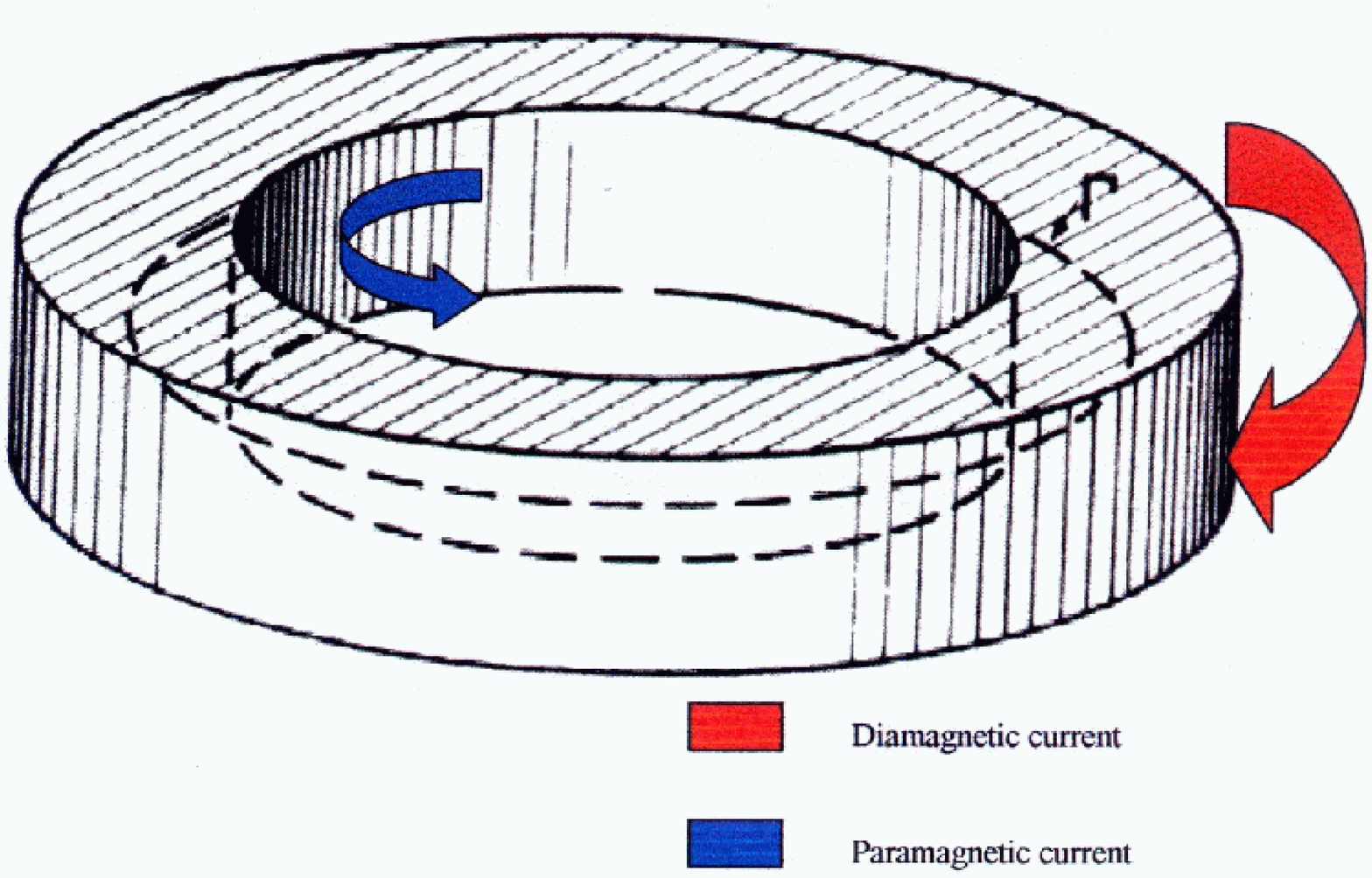,height=2.74in}}
\end{center}
\begin{center}
Figure 2: the path $\Gamma $
inside the superconductive ring
\end{center}
\end{figure}


We write this integral in the following way: 
\begin{equation}
\hbar \stackunder{\Gamma }{\oint }\nabla \theta \cdot d\overrightarrow{s}=q%
\stackunder{\Gamma }{\oint }\overrightarrow{A}\cdot d\overrightarrow{s}.
\label{int}
\end{equation}
Using the Stokes Theorem, equation (\ref{int}) becomes 
\begin{equation}
q\stackunder{\Gamma }{\oint }\overrightarrow{A}\cdot d\overrightarrow{s}=%
\stackunder{\Sigma }{\iint }curl\overrightarrow{A}\cdot d\overrightarrow{%
\sigma }.  \label{sto}
\end{equation}
From the laws of electromagnetism we have: 
\begin{equation}
\overrightarrow{B}=\nabla \wedge \overrightarrow{A}.  \label{b}
\end{equation}
Therefore, from equations (\ref{sto}) and (\ref{b}) we find, 
\begin{equation}
q\stackunder{\Gamma }{\oint }\overrightarrow{A}\cdot d\overrightarrow{s}=%
\stackunder{\Sigma }{\iint }\overrightarrow{B}\cdot d\overrightarrow{\sigma }%
=\Phi _{m}  \label{phy}
\end{equation}
$\Phi _{m}$ is the magnetic flux through the superconducting ring's hole.

Placing equation (\ref{phy}) into equation (\ref{int}) we get 
\begin{equation}
\hbar \stackunder{\Gamma }{\oint }\nabla \theta \cdot d\overrightarrow{s}=q%
\stackunder{\Sigma }{\iint }\overrightarrow{B}\cdot d\overrightarrow{\sigma }%
=q\Phi _{m}.  \label{nab}
\end{equation}
As the integral of the gradient of a function between two points, is equal
to the difference between the values that the function will assume at those
points we can write 
\begin{equation}
\stackunder{\Gamma }{\oint }\nabla \theta \cdot d\overrightarrow{s}=\theta
_{2}-\theta _{1}.  \label{clov30}
\end{equation}
What happens if the point (1) $\equiv $ point (2), i.e., when we go around a
closed path inside the ring? The value of $\theta $ after a complete
roundtrip must give the same value when inserted into the wave function, 
\begin{equation}
\psi \left( \overrightarrow{r}\right) =\sqrt{\rho \left( \overrightarrow{r}%
\right) }e^{i\theta \left( \overrightarrow{r}\right) }.  \label{clov31}
\end{equation}
This occurs if $\theta $ changes by $2\pi n$ with $n\in N$. Therefore, 
\begin{equation}
\stackunder{\Gamma }{\oint }\nabla \theta \cdot d\overrightarrow{s}=2\pi n.
\label{pi}
\end{equation}
Inserting equation (\ref{pi}) into equation (\ref{nab}) we see immediately
that the magnetic flux trapped inside the ring's hole will be quantized. 
\begin{equation}
n\frac{2\pi \hbar }{q}=\Phi _{m}  \label{flxoid}
\end{equation}
The magnetic flux quantum trapped inside the ring has the value: 
\begin{equation}
\Phi _{m_{0}}=\frac{h}{2e}.  \label{flxquanta}
\end{equation}
$q=2e$ because in a superconductor we have Cooper pairs as mentioned above.

Recognizing that in a SC, the supercurrent $\overrightarrow{j}$ is
proportional to the (ordinary) momentum (\ref{canon}) of Cooper pairs of
charge $2e$ and mass $2m$, we have 
\begin{equation}
n\frac{h}{2e}=n\Phi =\frac{m}{de^{2}}\stackunder{\Gamma }{\oint }%
\overrightarrow{j}\overrightarrow{ds}+\stackunder{\Sigma }{\iint }%
\overrightarrow{B}\cdot d\overrightarrow{\sigma }  \label{clov33}
\end{equation}
where $d$ in the denominator is the density of superconducting electrons,
while $\overrightarrow{ds}$ denotes a line integral around a closed contour
and $\overrightarrow{d\sigma }$ denotes a surface integral over a closed
contour. The contour $\Gamma $ is now (unlike in equation (\ref{nab})) a
general contour, and is not necessarily in the bulk of the SC where $%
\overrightarrow{j}=0\,$always. Equation (\ref{clov33}) defines the Magnetic
Fluxoid quantum $\Phi $ on the left hand side. The left hand side is called
the fluxoid rather than the flux because it is the sum of a flux term and
another term. Since the first term on the right hand side is a line
integral, and since the current $\overrightarrow{j}$ vanishes deep inside
the SC, for any contour in the bulk SC, this term can be neglected, and we
can say that the magnetic flux, defined by the second term right hand side,
is quantized.

Notice that it is the flux that is quantized, not the field.

Mathematically, the vanishing of the first term of (\ref{clov33}) makes
perfect sense. However, it is a little more difficult to interpret in term
of physical measurements. To do that, we will rewrite the first term by the
Stokes theorem to obtain 
\begin{equation}
n\frac{h}{2e}=n\Phi =\frac{m}{de^{2}}\stackunder{\Sigma }{\iint }\left(
\nabla \wedge \overrightarrow{j}\right) \overrightarrow{d\sigma }+%
\stackunder{\Sigma }{\iint }\overrightarrow{B}\cdot d\overrightarrow{\sigma }%
.  \label{clov34}
\end{equation}
The screening layer with the current is contained in the surface integral
when the contour is in the bulk SC, so we interpret (\ref{clov34}) to mean
that an experiment measures the TOTAL magnetic field due to the flux through
the hole (second term) PLUS\ the magnetic flux generated by the London
current layer around the hole (first term). (In a SC, $\nabla \wedge 
\overrightarrow{j}$ is proportional to the magnetic field by the London
Equations). Since that screening current is precisely the current necessary
to establish and maintain the flux through the hole at a quantized value
according to theory, experiments confirmed the theory when they detected a
quantum of flux given by the left hand side. Indeed, if the applied field
had a flux of, say, $(n+\frac{1}{4})\Phi $, a current is set up to move the
trapped flux in the hole to an even quantized value of $n\Phi $. There is
also a very small magnetic flux contribution from the penetration of the
external field into the London layer around the hole (from which it is not
screened, unlike the rest of the SC). This would show up as a slight
deviation from the quantized value.

If the SC is very thin, such that its thickness is no greater than the
penetration depth, effectively, there is no bulk SC, and the contour goes
right through the London layer around the hole. The interpretation of this
in light of (\ref{clov34}) is that the London current layer is no longer
inside the integration surface area and cannot contribute. Therefore, the
only term left is the second, with the result that a measurement now sees a
deficit in the flux, since the first term has effectively been excluded.
This deficit was also observed in experiments.

\chapter{\label{Q}Quantization of the Gravitomagnetic Flux in Superconductors
}

Not long after the detection of magnetic flux quantization in SC \cite
{Deaver}, physicists began to examine the role of GM in SC, recognizing the
analogy between the magnetic and GM vector potentials \cite{Dewitt}, \cite
{Papini}. Ross \cite{Ross} concluded that in the presence of a GM field,
there is not a pure Meissner effect inside a bulk SC. Li and Torr first
extended this result \cite{Li} and then proposed \cite{Li2} a novel GM
amplification mechanism arising from lattice ion contributions and
modification of the usual EM constitutive relations.

To include the contribution of the gravitomagnetic flux we need to write
down the Hamiltonian of an electrically charged particle with mass, moving
simultaneously in an electromagnetic, a gravitomagnetic, and gravitational
fields \cite{Ross}. 
\begin{equation}
\widehat{H}=\frac{1}{2m}\left( \frac{\hbar }{i}\nabla -q\overrightarrow{A}-4m%
\overrightarrow{A}_{g}\right) ^{2}+q\varphi -m\Phi .  \label{Ham}
\end{equation}
Inserting equation (\ref{Ham}) into the Schr\"{o}dinger equation $-\frac{%
\hbar }{i}\frac{\partial \psi }{\partial t}=\widehat{H}\psi $ we get: 
\begin{equation}
-\frac{\hbar }{i}\frac{\partial \psi }{\partial t}=\frac{1}{2m}\left( \frac{%
\hbar }{i}\nabla -q\overrightarrow{A}-4m\overrightarrow{A}_{g}\right)
^{2}\psi +q\varphi \psi -m\Phi \psi .  \label{s2}
\end{equation}
We see clearly from this equation that the Cooper pair canonical momentum
is: 
\begin{equation}
\overrightarrow{P}=\hbar \nabla \theta =m\overrightarrow{v}-q\overrightarrow{%
A}-4m\overrightarrow{A}_{g}.  \label{PP}
\end{equation}
The solution of equation (\ref{s2}) is the wave equation 
\begin{equation}
\psi \left( \overrightarrow{r}\right) =\sqrt{\rho \left( \overrightarrow{r}%
\right) }e^{i\theta \left( \overrightarrow{r}\right) }.  \label{W2}
\end{equation}
Notice that this equation can also be expressed as a function of the Cooper
pair mass density 
\begin{equation}
\psi \left( \overrightarrow{r}\right) =\frac{e}{m}\sqrt{\mu \left( 
\overrightarrow{r}\right) }e^{i\theta \left( \overrightarrow{r}\right) }.
\label{clov35}
\end{equation}
Placing (\ref{W2}) into (\ref{s2}) we will obtain the equations that govern
the dynamical behavior of $\rho $ and $\theta $. We have to keep in mind
that $\rho \left( x,y,z\right) $ and $\theta \left( x,y,z\right) $ are
functions of $x,y$ and $z$. Separating the real part from the imaginary part
we obtain the following couple of equations:

\begin{enumerate}
\item  The continuity equations for electric charge and mass govern the
behavior of $\rho $ and $\mu $%
\begin{equation}
\frac{\partial \rho }{\partial t}=\nabla \cdot \rho \overrightarrow{v}%
=\nabla \cdot \overrightarrow{j}\qquad and\qquad \frac{\partial \mu }{%
\partial t}=\nabla \cdot \mu \overrightarrow{v}=\nabla \cdot \overrightarrow{%
p}.  \label{cont}
\end{equation}

\item  The second equation governs the behavior of $\theta $. 
\begin{equation}
\hbar \frac{\partial \theta }{\partial t}=-\frac{1}{2}mv^{2}+q\varphi -m\Phi
-\frac{\hbar ^{2}}{2m}\left\{ \frac{1}{\sqrt{\rho }}\triangle \left( \sqrt{%
\rho }\right) \right\}   \label{sec}
\end{equation}
The right hand side of this equation is composed of the kinetic energy, the
electrical potential energy, the gravitational potential energy and an
additional term corresponding to an energy which has a ''quantum nature''
(we will take the density $\rho $ to be constant, so this fourth term will
vanish).
\end{enumerate}

In order to make the physics behind equation (\ref{sec}) more apparent we
can take the gradient of equation (\ref{sec}) and consider the canonical
momentum (\ref{PP}) that we obtained above. 
\begin{equation}
\frac{\partial \overrightarrow{v}}{\partial t}=\frac{q}{m}\left( -\nabla
\varphi -\frac{\partial \overrightarrow{A}}{\partial t}\right) +\left(
-\nabla \Phi -4\frac{\partial \overrightarrow{A}_{g}}{\partial t}\right) -%
\overrightarrow{v}\wedge \left( \nabla \wedge \overrightarrow{v}\right)
-\left( \overrightarrow{v}\cdot \nabla \right) \overrightarrow{v}
\label{prontinha}
\end{equation}
The first term on the right hand side of equation (\ref{prontinha})
corresponds to the electrical force, because: 
\begin{equation}
\overrightarrow{E}=-\nabla \varphi -\frac{\partial \overrightarrow{A}}{%
\partial t}.  \label{elec}
\end{equation}
The second term on the right hand side of equation (\ref{prontinha})
corresponds to the gravitational force, because as we saw above with
equation (\ref{equmo}), we have: 
\begin{equation}
\overrightarrow{g}-3\frac{\partial \overrightarrow{A}_{g}}{\partial t}=-%
\frac{\partial \Phi }{\partial x^{i}}-4\frac{\partial \overrightarrow{A}_{g}%
}{\partial t}  \label{peso}
\end{equation}
The third term on the right hand side of equation (\ref{prontinha})
corresponds to the Lorentz force and to the gravitational Lorentz force,
because, taking the curl of $\overrightarrow{v}$ that we extracted from
equation (\ref{PP}), we obtain: 
\begin{equation}
\nabla \wedge \overrightarrow{v}=-\frac{q}{m}\nabla \wedge \overrightarrow{A}%
-4\nabla \wedge \overrightarrow{A}_{g}=-\frac{q}{m}\overrightarrow{B}-4%
\overrightarrow{B}_{g}.  \label{Lor}
\end{equation}
Moving the fourth term on the right hand side of equation (\ref{prontinha})
to left hand side we obtain the comoving acceleration: 
\begin{equation}
m\left[ \frac{d\overrightarrow{v}}{dt}\right] _{Comoving}=\frac{\partial 
\overrightarrow{v}}{\partial t}+\left( \overrightarrow{v}\cdot \nabla
\right) \overrightarrow{v}  \label{com}
\end{equation}

Therefore, the equations of motion for an ideal electrically charged fluid
with mass which is located simultaneously in an electromagnetic, a
gravitational, and a gravitomagnetic field are: 
\begin{equation}
m\left[ \frac{d\overrightarrow{v}}{dt}\right] _{Comoving}=q\left( 
\overrightarrow{E}+\overrightarrow{v}\wedge \overrightarrow{B}\right)
+m\left( \overrightarrow{g}-3\frac{\partial \overrightarrow{A}_{g}}{\partial
t}+4\overrightarrow{v}\wedge \overrightarrow{B}_{g}\right)   \label{two}
\end{equation}
and 
\begin{equation}
\nabla \wedge \overrightarrow{v}=-\frac{q}{m}\overrightarrow{B}-4%
\overrightarrow{B}_{g}.  \label{three}
\end{equation}
From equation (\ref{three}) we know that the sum of the magnetic flux and
the GM flux is proportional to the circulation of the Cooper pair's velocity
around a closed path. We have indeed: 
\begin{equation}
\stackunder{\Sigma }{\int }\left( \nabla \wedge \overrightarrow{v}\right) d%
\overrightarrow{\sigma }=-\frac{q}{m}\stackunder{\Sigma }{\int }%
\overrightarrow{B}\cdot d\overrightarrow{\sigma }-4\stackunder{\Sigma }{\int 
}\overrightarrow{B}_{g}\cdot d\overrightarrow{\sigma }.  \label{ponto}
\end{equation}
Moreover, we see from the middle of equation (\ref{Lor}) that 
\begin{equation}
-\overrightarrow{v}=\frac{q}{m}\overrightarrow{A}+4\overrightarrow{A}_{g}.
\label{ve}
\end{equation}
The first term on the right hand side of equation (\ref{ve}), we interpret
to be the GM vector potential induced by the applied magnetic field \cite{Li}%
. Therefore, the Cooper pair velocity is proportional to the total
gravitomagnetic vector potential.

To include the contribution of GM in the magnetic fluxoid equation (\ref
{clov34}), one need simply add an appropriate term of $4(2m)\overrightarrow{B%
}_{g}$ in the right hand side of this equation, which becomes 
\begin{equation}
n\frac{h}{2e}=n\Phi =\frac{m}{de^{2}}\stackunder{\Sigma }{\iint }\left(
\nabla \wedge \overrightarrow{j}\right) \overrightarrow{d\sigma }+%
\stackunder{\Sigma }{\iint }\overrightarrow{B}\cdot d\overrightarrow{\sigma }%
+4\left( \frac{2m}{2e}\right) \stackunder{\Sigma }{\iint }\overrightarrow{B}%
_{g}\cdot d\overrightarrow{\sigma }.  \label{oid}
\end{equation}
Rearranging, we get 
\begin{equation}
n\frac{h}{2m}=n\Phi =\frac{1}{de}\stackunder{\Sigma }{\iint }\left( \nabla
\wedge \overrightarrow{j}\right) \overrightarrow{d\sigma }+\frac{e}{m}%
\stackunder{\Sigma }{\iint }\overrightarrow{B}\cdot d\overrightarrow{\sigma }%
+4\stackunder{\Sigma }{\iint }\overrightarrow{B}_{g}\cdot d\overrightarrow{%
\sigma }  \label{coco}
\end{equation}
We have transformed the quantization relation into a form for which the left
hand side has dimensions of GM-flux ($m^{2}/s$). A single GM flux quantum
trapped inside the ring has the value: 
\begin{equation}
\Phi _{g_{0}}=\frac{h}{2m}.  \label{GMF}
\end{equation}
We interpret the right hand side as being the total GM flux in the system,
which is quantized. The first term is that GM flux from the moving currents
in the London layer; the third term is the ''pure'' GM\ flux from the
external field; the second term we interpret to be the GM flux induced by
the applied magnetic field. Note that the GM quantum number is the same as
the magnetic one for the same physical situation. Quantization of one
establishes the other.

Since the first term is relatively small, the important question is whether
the interpretation of the second term is valid. This second term will not be
small in general, if for no other reason than the factor of $1/m$ is so
large. Traditional interpretations of the flux quantization relations
emphasize that what is physically measured in an experiment is only one term
of those relations, say, the magnetic flux integral. Each term is measured
separately. The left hand side, that which is actually quantized, is not
measurable and just represents a mathematical agglomeration of the
individual terms on the right hand side. Support for this interpretation
comes from experiments \cite{Tilley}, \cite{Rose} on very thin SC with
thickness less than the London penetration depth. Unlike bulk SC, thin SC do
not exhibit magnetic flux quantization. Explanation for this is offered by
observing that the current line integral in equation (\ref{clov33}) is
non-vanishing for thin SC because the contour can not be placed in the bulk
where the Meissener Effect prevails. Since this term does not vanish, and
the left hand side of equation (\ref{clov33}) is fixed, this must mean the
magnetic flux integral is less than the flux quantum, which accounts for the
measurements. This is certainly mathematically sensible and persuasive.

However, as we saw in equation (\ref{clov34}), we can write this equation in
a form where the current integral is also written as a flux integral, so
that all terms have the same form. Now, this term will not automatically
vanish in all cases, since $\nabla \wedge \overrightarrow{j}$ is
proportional to the magnetic field in a SC by one of the London Equations.
And the surface of integration includes that region containing the
paramagnetic current, where a magnetic field will exist. Then, a more
physically natural and persuasive explanation for this term is that it
represents the magnetic flux contribution of the paramagnetic current, which
generates the magnetic field sufficient to maintain a quantized flux. It is
the left hand side of the equation which is actually measured, representing
the combined contributions of the external magnetic field, a contribution
from any induced magnetic fields (e.g. the negligible external GM-induced
magnetic field), and the paramagnetic field. In a bulk SC, these
contributions sum to ensure the measured magnetic flux is quantized.

In a very thin SC, the integration surface is so small, it chops off part of
the paramagnetic region. There is simply not enough SC left to maintain the
entire quantized flux; it is not a complete SC system. This explains why the
total flux from all the sources on the right hand side is insufficient to
impose the quantized value.

A corresponding interpretation of the GM flux quantization relation equation
(\ref{coco}) implies that what would be measured in a GM experiment is the
quantized GM flux value on the left hand side of equation (\ref{coco}),
which is entirely of GM nature. That is, the GM flux Quantum represents a
physically observable quantity, just as the magnetic flux quantum does.
Quantization of the one imposes quantization on the other. As in the purely
magnetic case, a sufficiently large external magnetic field must be present
to boost the SC into a $n>0$ quantum state. Though the Earth's GM field
pervades the terrestrial environment, it is not strong enough to do this.
for this reason, quantized GM flux due to an external GM field alone will
not be observable. But because the magnetic and GM fluxes are tied together
by the same quantum number in the same quantum state, if a sufficiently
large external magnetic field is present, it should be possible to observe
its GM counterpart. Indeed, the proposed experiment may offer the first
direct test of the thin SC explanation by searching for different fields
representing the multiple terms on the right hand side of the flux equations.

An objection, both cogent and obvious, can be made to this interpretation on
the grounds that there is no known mechanism for producing a macroscopically
measurable GM field. Exotic mechanisms can be imagined. Coherent zero point
motion was suggested for a similar purpose in the context of superfluid HeII 
\cite{Becker}. It turns out this was not the first time Coherent zero point
motion was proposed as an underlying mechanism operative in superfluids.
Twenty years ago, Shenoy and Biswas \cite{Shenoy} proved by detailed
statistical mechanical arguments using a coherent state formalism in
coordinate space that coherent, macroscopic zero point motion characterizes
and defines an interacting Bose gas (that is, a non-ideal Bose gas)
superfluid system. It is only ideal Bose gases that display Off-Diagonal
Long-Range Order (ODLRO) and form a true condensate; real superfluids do not
possess ODLRO. Instead, any interaction effectively mediates a form of phase
coherence which is reflected in a momentum space dispersion. coherent zero
point motion is at first glance counterintuitive, if not downright
oxymoronic: Zero point motion is, after all, supposed to be inherently
random. A very schematic way of thinking about this physically in terms of
random phase space orbits of the Bose gas constituents, which are
nonetheless correlated in position along each such orbit, has been suggested 
\cite{Becker}.

Outlandish as some of these ideas may appear, there are some precedents for
the role of ''background'' contributions in condensed-matter systems.
Certain thin film superfluid helium mixture systems \cite{Hallock} have
excited states, each of which separates out into completely independent
quantum subsystems. Extensive quantities are determined by simply summing
the contribution of each individual state. The Fractional Quantum Hall
Effect can be understood in terms of a system of anyonic, fractionally
charged quasi-particle excitations coupled to magnetic flux quanta \cite
{Frohlich}. These collective excitations are ''hierarchically'' correlated
and themselves form a condensate \cite{Frohlich}, \cite{Becker}. Thus, it is
not entirely inconceivable that an amplification effect from repeated
''reuse'' of hierarchies of underlying collective coherent constituents
might begin to explain the predicted macroscopic GM flux quantum. Pushing
such speculation to the limits, one may even posit that such interactions
may explain the extraordinary coherence distances of recent entangled
quantum state teleportation experiments \cite{Furusawa}.

So the experimental method, presented below, hinges on that second term: if
it does not physically contribute to a GM flux, but just represents the
magnetic torque that a magnetized body would feel if such a body were
present in the SC hole, then the other ''pure'' GM term is far too small to
be measurable. On the other hand, if the two terms of equation (\ref{coco})
and the fluxoid itself are GM in origin, the effect may be large and
definitely noticeable given the right circumstances.

\chapter{Gravitomagnetic Flux Experiment}

\section{Gravitomagnetic Flux Experiment Concept}

The actual experiment concept \cite{Beckart} consists of suspending a SC
from a balancing-torsion mechanism akin to that used in the Cavendish
experiment (see figure 3). A hole is scooped out of the bulk SC, such that a
small, rotating cylinder (C) of non-magnetic material can be placed inside.
The cylinder has its long axis oriented in the horizontal plane, and spins
along that axis.

The cylinder is supported by a mechanism which allows rotation around the
long axis, but resists rotation around any other. The SC itself is free to
rotate in the horizontal plane. There is a small mirror (M1) or other such
rotation sensor on the SC suspension assembly.

If a strong enough total GM fluxoid quantum is present in the hole, due to a
large applied magnetic field, the cylinder ought to precess about an axis
perpendicular to the horizontal plane. However, the cylinder is constrained
from precessing. Since the SC, to which the cylinder should be coupled by
virtue of the GM field, is free to rotate in the precession plane, the
cylinder's precessional angular momentum is transferred to the SC. The SC
then twist about its suspension, permitting a measurement (R1) of the degree
of turn from the sensor on the suspension. Alternatively, a direct
measurement of the cylinder's precession can be attempted.


\begin{figure}
\begin{center}
\mbox{\psfig{file=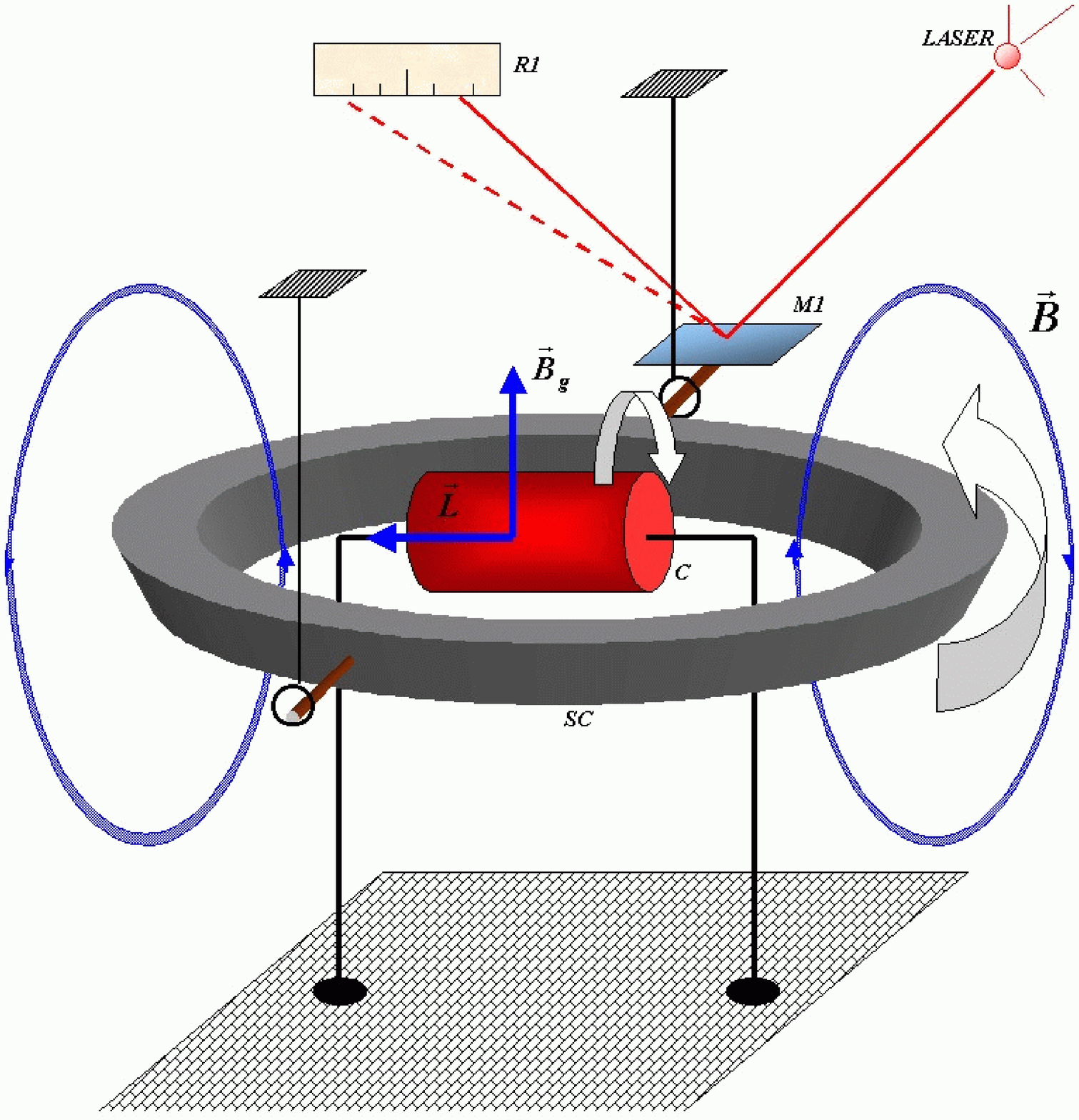,height=6.58in}}
\end{center}
\begin{center}
Figure 3: The Gravitomagnetic Cavendish balance
\end{center}
\end{figure}


To calibrate the apparatus and ensure that there are no lingering magnetic
effects in the cylinder, the cylinder set-up experiment should first be run
with no SC present, with just the applied field. Next, the experiment should
be repeated with a SC, but at $T>T_{c}$, where there should be no
quantization effects. Finally, the full SC plus cylinder tests can be run,
varying the applied magnetic field as desired, as well as other parameters.
A magnetized cylinder should be substituted for the non-magnetic test body,
not only to verify the basic experimental set-up in case no GM effect is
seen, but to also verify that an experimental effect is present even in the
well-known case of the SC magnetic field.

Flexibility in the experiment design should be sufficient to permit
variations in the following experimental parameters:

\begin{enumerate}
\item  The axis of the SC can be moved smoothly and without induced
vibrations in, or causing collisions between, the rotating cylinder and the
SC. This can be done under cryogenic and vacuum conditions.

\item  The speed and direction of the cylinder rotation can be changed by
varying the flow rates of the helium or hydrogen jets directed at the top
and bottom of the cylinder, under cryogenic and vacuum conditions.

\item  The magnitude of the trapped magnetic flux can be adjusted after each
run of the experiment sequence by changing the applied magnetic field.

\item  The composition of the cylinder and its mass properties (e.g. moment
of inertia) are adjustable by switching out cylinder samples.
\end{enumerate}

The following tables illustrate schematically the way parameter space can be
scanned to test whether there is or is not a macroscopic quantized GM flux
separate and distinguishable from magnetic flux. In Table 2, it is assumed
there is an applied magnetic field equivalent to some integer multiple of
the magnetic flux quantum. In Table 3, it is assumed that some non-integer
(say, $n+1/4$) quantum of magnetic flux is applied. Each entry in the tables
contains an ''equivalent'' magnetic (for a magnetic test body) or GM (for a
non-magnetic test body) flux strength that would be left by the test body as
a torque, in the different temperature regimes. If there is no macroscopic
GM flux in the superconducting regime, then all entries in the second column
would be zero in both tables.

\begin{center}
\begin{tabular}{|c|c|c|}
\hline
& Magnetic Test Body & Non-magnetic Test Body \\ \hline
$T<T_{c}$ & $n$ & $n$ \\ \hline
$T>T_{c}$ & $n$ & $0$ \\ \hline
\end{tabular}

Table 2: Integer Quantized Applied Magnetic Flux

\bigskip 

\begin{tabular}{|c|c|c|}
\hline
& Magnetic Test Body & Non-magnetic Test Body \\ \hline
$T<T_{c}$ & $n$ & $n$ \\ \hline
$T>T_{c}$ & $n+1/4$ & $0$ \\ \hline
\end{tabular}

Table3: Non-integer Quantized Applied Magnetic Flux
\end{center}

\section{Gravitomagnetic Force Calculation}

In this section is outlined the calculations needed to compute the forces
that are expected to arise from the GM-induced precession of the cylinder
based on the theory of section 4.

For a given applied magnetic field, the fluxoid quantum number is obtained
from 
\begin{equation}
n=\pi r^{2}H\frac{2e}{h}  \label{n}
\end{equation}
where $H$ is the applied magnetic field, $r$ is the radius of the hole.
Since n is an integral quantum number, it should be rounded to the nearest
integer.

Since it is believed that the GM field can penetrate the SC, unlike the
magnetic field, the area that enters the GM fluxoid is that of the whole SC,
not the hole. From the left hand side and the the third term on the right
hand side of equation (\ref{coco}) we estimate the GM field as an average
over its penetration area. This will just be an estimate because, at least
in the case of the magnetic field, the field is not uniform in a flat
cylindrical SC as will be used for this experiment. Conversely, if the GM
field is not expelled from the bulk SC, its lines should not deviate from
linearity to any degree, making the average field estimate from the GM flux
a better estimate than one for $\overrightarrow{B}$ itself, 
\begin{equation}
B_{g}=n\frac{h}{8m\pi R^{2}}=\frac{1}{4}\frac{e}{m}H\left( \frac{r}{R}%
\right) ^{2}  \label{x1}
\end{equation}
where $R$ is the radius of the SC.

If $B_{g}$ is given by (\ref{x1}), the precession velocity of the cylinder
is therefore 
\begin{equation}
\Omega =2B_{g}=\frac{1}{2}\frac{e}{m}H\left( \frac{r}{R}\right) ^{2}.
\label{x2}
\end{equation}
The torque the GM field will attempt to apply to the cylinder so as to
result in the precession velocity of the cylinder is therefore 
\begin{equation}
\overrightarrow{N}=2\overrightarrow{L}\Lambda \overrightarrow{B}_{g}=\frac{e%
}{2m}\left( \frac{r}{R}\right) ^{2}\overrightarrow{L}\wedge \overrightarrow{H%
}.  \label{x3}
\end{equation}
Rigid body spin angular momentum is given by 
\begin{equation}
\overrightarrow{L}=I\overrightarrow{\omega }  \label{x4}
\end{equation}
where $\overrightarrow{\omega }$ is the spin angular velocity. A cylinder
rotating around its long axis has moment of inetia $I$ of 
\begin{equation}
I=\frac{M\alpha ^{2}}{2}  \label{x5}
\end{equation}
where $M$ is the mass of the cylinder and $\alpha $ is its radius. Putting
both (\ref{x4}) and (\ref{x5}) into (\ref{x3}), we obtain (assuming right
angles) 
\begin{equation}
N=\frac{M\alpha ^{2}e}{4m}\left( \frac{r}{R}\right) ^{2}\omega H.  \label{x6}
\end{equation}
Since the cylinder will be fixed in the precessional rotation direction, but
the SC will be free, the back coupling of the cylinder to SC should transfer
the applied torque to the SC, resulting in a torque, and therefore force, on
the SC suspension mechanism. The torque is given by 
\begin{equation}
\overrightarrow{N}=\overrightarrow{R}\wedge \overrightarrow{F}.  \label{x7}
\end{equation}
So, assuming right angles relative to the apparatus, 
\begin{equation}
F=\frac{N}{R}=\frac{M\alpha ^{2}e}{4m}\frac{r^{2}}{R^{3}}\omega H.
\label{x8}
\end{equation}

For the following suggested parameter values

\begin{itemize}
\item  $M$, Mass of test body: $2\times 10^{-3}$ $Kg$

\item  $\omega $, Angular velocity of test body: $10$ $Rad/s$

\item  $R$, Radius of SC: $1.1\times 10^{-2}$ $m$

\item  $r$, Radius of SC\ hole: $5\times 10^{-3}$ $m$

\item  $\alpha $, Radius of test body: $3\times 10^{-4}$ $m$

\item  $m$, electron mass: $9.1091\times 10^{-31}$ $Kg$

\item  $e$, electron charge: $1.6021\times 10^{-19}$ $C$

\item  $h$, Planck constant: $6.6256\times 10^{-34}$ $J.s$
\end{itemize}

the value of $n$ at the lower critical field ($2\times 10^{-2}$ $T$) for the
material proposed for this experiment, with $r$ as above, is from (\ref{n}) 
\[
n=7.59\times 10^{8}. 
\]
This would make the force 
\[
F=29.7\text{ }Newtons! 
\]
For a more reasonable magnetic field, like the earth's ($3\times 10^{-5}$ $T$%
) we get 
\[
n=1.139\times 10^{6} 
\]
and the force is 
\[
F=4.46\times 10^{-2}\text{ }Newton. 
\]
It is essential to use small-area holes in the SC. While it is the flux
which is quantized, it is the field which induces the precession. Too large
a hole means that the flux can be dominated by the hole area, rather than
the field, with the result that precessional effects could be quite small,
despite a large applied magnetic field. Conversely, it is not quite
customary to suspend small test bodies in small holes in SC, which may
explain why these effects have not been noticed before.

\chapter{Conclusion}

For suitable superconductors geometries, in a sufficiently strong external
magnetic field below the superconducting transitions temperature, measurable
GM fields should be detected. Such GM fields obey a GM version of the
familiar magnetic fluxoid relation, defining a GM Flux quantum. This GM Flux
quantum, consistent with our reinterpretation of the magnetic fluxoid
relation, has real physical significance. It is not the outcome of purely
arbitrary algebraic manipulations of the magnetic fluxoid relation; nor is
it an unmeasurable quantity which is merely the arithmetic sum of physical
quantities which can only be measured separately. Rather, the flux quanta
represent the total contribution of the physical sources of the fields
encompassed by each quanta, each source ultimately traceable to the
Hamiltonian obeyed by the condensate system. Both the magnetic and GM fields
are attributed to the same carriers (in the present conception of the
theory), namely, the Cooper pairs, and so must be governed by the same
quantum number. Since terrestrial GM fields themselves are too weak to
promote a superconductor out of the flux quantum ground state, reliance is
placed on magnetic fields to boost the superconductor out of its ground
state, creating the same number of non-vanishing flux quanta for both the
magnetic and GM fields.

Our experiment concept takes advantage of this magnetic boost to detect GM
fields ''trapped'' in a superconductor. While it may be preferable to
perform this experiment with low-temperature superconductor systems because
of their relative simplicity, high-Tc superconductors should also be
amenable to this investigation. Indeed, while this concept is not without
its own complexities, a proof of principle demonstration, using larger
magnetic fields, should be possible with standard superconductor laboratory
equipment supplemented by the Cavendish-type apparatus. If counting quanta
is perhaps the most fundamental way of assessing the relative magnitude of a
field, then such a detection also brings us into the ''mesoscopic'' regime
of gravitational fields, featuring $10^{2}-10^{4}$ GM flux quanta.

Such demonstration would hold profound implications for the nature of the
gravitational field, its quantization, its relation to other classical
fields, its relation to coherence, the transition from linearized to
nonlinear regimes of field equations, and the relationship between gauge
fields, their field equations, and their topologies. Other systems, such as
superfluids and the exotic coupled superfluid-superconductor systems thought
to exist in neutron stars, should also manifest quantized GM effects to
varying degrees. Practical applications, if the effects are large enough,
may also be feasible (generation of gravitational fields through time
varying GM flux).

Even if no macroscopic GM field is found, this experiment still represents
an important verification of the traditional fluxoid interpretation. There
may be little doubt harbored in that interpretation after several decades of
experimentation with thin superconductors, interferometric phase and
Josephson Effect measurements. Yet, this experiment, perhaps for the first
time, tests that interpretation by seeking direct, physical effects,
distinguishable by the presence of the different fields that produce them,
on objects of distinctive composition. A phase shift is a phase shift no
matter how produced. But a GM field and a magnetic field should have quite
different effects on objects composed of different materials.

A deeper, completly explanatory theory for this effect trails far behind the
basic concept at present. Halting steps towards a deeper theory, invoking
other speculative ideas such as coherent zero point motion, have been taken.
But a complete description may have to await discovery of the effect and
fuller exploration of its intricacies and ramifications.

\end{document}